\documentclass[12pt,a4paper,final]{revtex4}

\usepackage{sidecap}
\usepackage{ulem}
\usepackage{epsfig}
\usepackage{amsmath,amssymb,amsthm}
\usepackage{graphicx}
\usepackage{bm}
\usepackage{color}

\usepackage{float}

\DeclareMathAlphabet{\mathpzc}{OT1}{pzc}{m}{it}

\begin{document}

\renewcommand{\textfraction}{0.00}


\newcommand{\vAi}{{\cal A}_{i_1\cdots i_n}} 
\newcommand{\vAim}{{\cal A}_{i_1\cdots i_{n-1}}} 
\newcommand{\vAbi}{\bar{\cal A}^{i_1\cdots i_n}}
\newcommand{\vAbim}{\bar{\cal A}^{i_1\cdots i_{n-1}}}
\newcommand{\htS}{\hat{S}} 
\newcommand{\htR}{\hat{R}}
\newcommand{\htB}{\hat{B}} 
\newcommand{\htD}{\hat{D}}
\newcommand{\htV}{\hat{V}} 
\newcommand{\cT}{{\cal T}} 
\newcommand{\cM}{{\cal M}} 
\newcommand{\cMs}{{\cal M}^*}
\newcommand{\vk}{\vec{\mathbf{k}}}
\newcommand{\bk}{\bm{k}}
\newcommand{\kt}{\bm{k}_\perp}
\newcommand{\kp}{k_\perp}
\newcommand{\km}{k_\mathrm{max}}
\newcommand{\vl}{\vec{\mathbf{l}}}
\newcommand{\bl}{\bm{l}}
\newcommand{\bK}{\bm{K}} 
\newcommand{\bb}{\bm{b}} 
\newcommand{\qm}{q_\mathrm{max}}
\newcommand{\vp}{\vec{\mathbf{p}}}
\newcommand{\bp}{\bm{p}} 
\newcommand{\vq}{\vec{\mathbf{q}}}
\newcommand{\bq}{\bm{q}} 
\newcommand{\qt}{\bm{q}_\perp}
\newcommand{\qp}{q_\perp}
\newcommand{\bQ}{\bm{Q}}
\newcommand{\vx}{\vec{\mathbf{x}}}
\newcommand{\bx}{\bm{x}}
\newcommand{\tr}{{{\rm Tr\,}}} 
\newcommand{\bc}{\textcolor{blue}}

\newcommand{\beq}{\begin{equation}}
\newcommand{\eeq}[1]{\label{#1} \end{equation}} 
\newcommand{\ee}{\end{equation}}
\newcommand{\bea}{\begin{eqnarray}} 
\newcommand{\eea}{\end{eqnarray}}
\newcommand{\beqar}{\begin{eqnarray}} 
\newcommand{\eeqar}[1]{\label{#1}\end{eqnarray}}
 
\newcommand{\half}{{\textstyle\frac{1}{2}}} 
\newcommand{\ben}{\begin{enumerate}} 
\newcommand{\een}{\end{enumerate}}
\newcommand{\bit}{\begin{itemize}} 
\newcommand{\eit}{\end{itemize}}
\newcommand{\ec}{\end{center}}
\newcommand{\bra}[1]{\langle {#1}|}
\newcommand{\ket}[1]{|{#1}\rangle}
\newcommand{\norm}[2]{\langle{#1}|{#2}\rangle}
\newcommand{\brac}[3]{\langle{#1}|{#2}|{#3}\rangle} 
\newcommand{\hilb}{{\cal H}} 
\newcommand{\pleft}{\stackrel{\leftarrow}{\partial}}
\newcommand{\pright}{\stackrel{\rightarrow}{\partial}}

\newcommand{\squeezeup}{\vspace{-2.5mm}}


\title{Importance of different energy loss effects in jet suppression at RHIC and LHC}

\date{\today}
 
\author{Bojana Blagojevic}

\author{Magdalena Djordjevic}
\affiliation{Institute of Physics Belgrade, University of Belgrade, Serbia}

\begin{abstract} 
Jet suppression is considered to be an excellent probe of QCD matter created in ultra-relativistic heavy ion collisions. Our theoretical predictions of jet suppression, which are based on our recently developed dynamical energy loss formalism, show a robust agreement with various experimental data, which spans across different probes, experiments (RHIC and LHC) and experimental conditions (i.e. all available centrality regions). This formalism includes several key ingredients, such as inclusion of dynamical scattering centers, finite size QCD medium, collisional energy loss, finite magnetic mass and running coupling. While these effects have to be included based on theoretical grounds, it is currently unclear what is their individual importance in accurately interpreting the experimental data, in particular because other approaches to suppression predictions commonly neglect some -  or all -  of these effects. To address this question, we here study the relative importance of these effects in obtaining accurate suppression predictions for D mesons (clear energy loss probe) at top RHIC and LHC energies. We obtain that several different ingredients are responsible for the accurate predictions, i.e. the robust agreement with the data is a cumulative effect of all the ingredients, though inclusion of the dynamical scattering centers has the largest relative importance. 
	
\end{abstract}

\maketitle 
\section{Introduction} 
Suppression of high transverse momentum light and heavy  flavor observables~\cite{Bjorken} is considered to be an excellent probe of QCD matter created in ultra-relativistic heavy ion collisions at RHIC and the LHC. One of the major goals of these experiments is mapping the QGP properties, which requires comparing available suppression data with the theoretical predictions~\cite{STE, STE1, STE2}. Such comparison tests different theoretical models and provides an insight into the underlying QGP physics. It is generally considered that the crucial ingredient for the reliable suppression predictions is an accurate energy loss calculation.

Therefore, we previously developed the dynamical energy loss formalism, which includes the following effects: \textit{i)} dynamical scattering centers, \textit{ii)} QCD medium of a finite size~\cite{DynEL, DynEL1}, \textit{iii)} both radiative~\cite{DynEL, DynEL1} and collisional~\cite{MD_Coll} energy losses, \textit{iv)} finite magnetic mass~\cite{MagM} and \textit{v)} running coupling~\cite{RunnC}. This energy loss formalism is based on the pQCD calculations in finite size and optically thin dynamical QCD medium, and has been incorporated into a numerical procedure~\cite{RunnC} that allows generating state-of-the art suppression predictions. 

These predictions are able to explain heavy flavor puzzle (the fact that, contrary to pQCD expectations, both light and heavy flavor probes have very similar experimentally measured $R_{AA}$) at both RHIC~\cite{CRHIC} and the LHC~\cite{HFLHC} and, in general, show a very good agreement with the available suppression data at these experiments, for a diverse set of probes~\cite{RunnC, CRHIC} and centrality regions~\cite{NCLHC}.

Such a good agreement of the predictions with the experimental data however raises a question of which energy loss effects are responsible for the accurate predictions. In other words, is there a single dominant energy loss effect which is responsible for the good agreement, or is this agreement the result of a superposition of several smaller improvements? This issue is moreover important, given the fact that various pQCD approaches~\cite{BDMPS,BDMS,Z,ASW,ASW1,AMY,GLV,DGLV,HT,HT1} to the energy loss calculations neglect some (or many) of these effects. 

Consequently, we here address the importance of different energy loss ingredients in the suppression calculations. For this purpose, it would be optimal to have a probe that is sensitive only to the energy loss, i.e. for which fragmentation and decay functions do not play a role. The D meson suppression is such a probe, since the fragmentation functions do not modify bare charm quark suppression, as previously shown in~\cite{CRHIC,HFLHC}. To explore different energy loss approximations, which have been used in suppression predictions, we here concentrate on the D meson suppression in central 200 GeV Au+Au collisions at RHIC and 2.76 TeV Pb+Pb collisions at the LHC. While high momentum D meson suppression data are not available at RHIC - the RHIC measurements extend up to 6 GeV - such data are available at the LHC, which is useful as a baseline for assessing the importance of different effects.
 
Our approach is to systematically include different energy loss effects. In particular, we first compare the relative importance of radiative and collisional contribution to the D meson suppression predictions, to assess the adequacy of the historically widely used static approximation. We then investigate the importance of including the dynamical scattering centers, followed by the collisional energy loss and the finite size (LPM) effect. Finally, we also address the importance of including  the finite magnetic mass and the running coupling. 

\section{Theoretical and computational frameworks}
In this section we first provide a brief overview of the computational framework and our dynamical energy loss formalism.  As mentioned above this formalism leads to a very good agreement with the suppression experimental data, across different probes, collision energies and centrality regions~\cite{RunnC, CRHIC, NCLHC}. We will also introduce how the energy loss expression is modified, as different ingredients are excluded from this formalism. Note that, in the Results and Discussion section, we will for clarity address different energy loss effects in a reverse order: i.e. we will start from the static approximation, and will systematically include all the effects, as such historically-driven approach is more comprehensible and easier to follow.   

For studying the importance of different energy loss effects, we will use angular averaged nuclear modification factor $R_{AA}$, which is well established as a sensitive observable for interaction of high-momentum particles with the QCD medium. The nuclear modification factor $R_{AA}$ is defined as the ratio of the quenched $A+A$ spectrum to the $p+p$ spectrum, scaled by the number of binary collisions $N_{\mathrm{bin}}$:
\begin{eqnarray}
R_{AA}(p_T)=\dfrac{dN_{AA}/dp_T}{N_{\mathrm{bin}} dN_{pp}/dp_T}
\label{RAA}\ .
\end{eqnarray}
Furthermore, since angular averaged $R_{AA}$ was previously shown to be sensitive almost entirely to the average properties (temperature) of the evolving medium (in distinction to elliptic flow, $v_2$, which is considered highly sensitive to the details of the medium evolution)~\cite{Thorsten,Molnar}, angular averaged $R_{AA}$ can be taken as a “nearly pure” test of the jet-medium interactions. Due to this, we do not consider the effects of the medium evolution in this study, but provide a detailed study of the importance of different jet-medium effects. For this purpose, we model the medium by assuming an effective temperature of 304 MeV at the LHC (as extracted by ALICE~\cite{LHC_T}) and effective temperature of 221 MeV  at RHIC (as extracted by PHENIX~\cite{RHIC_T}).

In order to obtain the quenched spectra, we use generic pQCD convolution~\cite{RunnC,PLF}:
\begin{eqnarray}
\frac{E_f d^3\sigma}{dp^3_f} = \frac{E_i d^3\sigma(Q)}{dp^3_i} \otimes P(E_i \rightarrow E_f)\ . 
\label{pQCD}
\end{eqnarray}
 
In Eq.~\eqref{pQCD} Q stands for charm quarks and $\dfrac{E_i d^3\sigma(Q)}{dp^3_i}$ denotes the initial charm quark spectrum computed at next-to-leading order~\cite{Heavy}. $P(E_i \rightarrow E_f)$ is the energy loss probability, which includes both radiative and collisional energy losses in a finite size dynamical QCD medium, multi-gluon~\cite{MGF} and path length~\cite{PLF, PLF1} fluctuations. The path length distributions are extracted from~\cite{PLF1}. In distinction to Eq.(1) from~\cite{RunnC}, in our calculations we do not use the fragmentation function $D(Q \rightarrow H_Q)$ of charm quark into D meson ($H_Q$), because fragmentation does not alter bare charm quark suppression~\cite{CRHIC,HFLHC}, nor do we use decay function $f(H_Q \rightarrow e)$, because D mesons are directly measured in the experiments. 

The expression for the radiative energy loss in a finite size dynamical QCD medium~\cite{DynEL, DynEL1}, obtained from HTL approximation, at $1^{st}$ order in opacity is given by:
\begin{eqnarray}
\frac{\Delta E_{rad}}{E} = \frac{C_R {\alpha_S}}{\pi} \frac{L}{\lambda}\int dx \frac{d^2k}{\pi} \frac{d^2q}{\pi} \,v(\mathbf{q})  \left(1{-}\frac{\sin \frac{(\mathbf{k}{+}\mathbf{q})^2{+}\chi}{x E^+}L}{\frac{(\mathbf{k}{+}\mathbf{q})^2{+}\chi}{x E^+} L}\right) 
  \frac{2 (\mathbf{k}{+}\mathbf{q})}{(\mathbf{k}{+}\mathbf{q})^2{+}\chi} \left(\frac{(\mathbf{k}{+}\mathbf{q})}{(\mathbf{k}{+}\mathbf{q})^2{+}\chi}{-}\frac{\mathbf{k}}{{\mathbf{k}}^2{+}\chi}\right).
\label{eloss}  
\end{eqnarray}

In Eq.~\eqref{eloss}, $v(\textbf{q})$ denotes the effective cross section defined below, L is the length of the finite size QCD medium, E is the jet energy, $\mathbf{k}$ is the transverse momentum of the radiated gluon, while $\mathbf{q}$ is the transverse momentum of the exchanged (virtual) gluon and $x$ represents the longitudinal momentum fraction of the jet carried away by the emitted gluon. The color factor is $C_R=\frac{4}{3}$. $\chi=M^2_c x^2 + m^2_g$, where $m_g=\mu_E/\sqrt{2}$ is the effective (asymptotic) mass for gluon with the hard momenta $k \gtrsim T$, while $\mu_E$ is Debye (electric) screening mass and $M_c=1.2$ GeV is the charm quark mass. ${\lambda}$ is the mean free path in the QCD medium and in the dynamical case is given by $\frac{1}{\lambda_{dyn}}= 3 {\alpha}_S T$. In the incoherent limit~\cite{DynEL}, 
$\dfrac{\sin \frac{(\mathbf{k}{+}\mathbf{q})^2{+}\chi}{x E^+}L}{\frac{(\mathbf{k}{+}\mathbf{q})^2{+}\chi}{x E^+} L} \rightarrow 0$.

The effective cross section, with the included finite magnetic mass~\cite{MagM}, is given by the equation below, where $\mu_M$ is the magnetic screening mass: 
\begin{eqnarray}
v(\mathbf{q})=\frac{{\mu^2_E}-{\mu^2_M}}{({\mathbf{q}}^2 + {\mu^2_E}) ({\mathbf{q}}^2+ {\mu^2_M})} \ .
\label{eqv}
\end{eqnarray}

Note that, in the case when magnetic mass is equal to zero, the above expression reduces to a well-known HTL effective cross section~\cite{AMY,DynEL}:
\begin{eqnarray}
v(\mathbf{q})=\frac{{\mu^2_E}}{{\mathbf{q}}^2({\mathbf{q}}^2 + {\mu^2_E}) } \ . 
\label{magnmass0}
\end{eqnarray}

Non-perturbative approaches~\cite{xb,xb4, xb1, xb2, xb3} suggest that at RHIC and the LHC the range of magnetic to electric mass ratio is $0.4 < \mu_M/\mu_E < 0.6$. We therefore use these values in Eq.~(\ref{eqv}), when generating suppression predictions in the case of the finite magnetic mass. In the case of zero magnetic mass, we use Eq.~(\ref{magnmass0}) above.

The collisional energy loss is calculated in accordance with~\cite{MD_Coll}, 
 i.e. we use Eq. (14) from that reference for the finite size QCD medium and Eq. (16) for the incoherent limit.

The running coupling is introduced according to~\cite{RunnC} and is defined as in~\cite{Rc}:
\begin{eqnarray}
\alpha_S(Q^2)=\frac{4 \pi}{(11-2/3 n_f) \ln(Q^2/\Lambda^2_{QCD})} \ ,
\label{alpha}
\end{eqnarray}
where $\Lambda_{QCD}$ is the perturbative QCD scale ($\Lambda_{QCD}=0.2$ GeV) and $n_f=2.5$ ($n_f=3$) for RHIC (LHC) is the number of the effective light quark flavors. In the case of the running coupling, Debye mass $\mu_E$~\cite{mju} is obtained by self-consistently solving the equation:
\begin{eqnarray}
\frac{\mu^2_E}{{\Lambda}^2_{QCD}} \ln\left(\frac{\mu^2_E}{{\Lambda}^2_{QCD}}\right)=\frac{1+n_f/6}{11-2/3 n_f} \left(\frac{4\pi T}{{\Lambda}_{QCD}}\right)^2 \ .
\label{mu}
\end{eqnarray}
Otherwise, when the running coupling is not included, fixed values of the strong coupling constant $\alpha_S=\frac{g^2}{4 \pi}=0.3$ for RHIC ($\alpha_S=0.25$ for LHC)~\cite{Betz} and Debye mass $\mu_E = g T$ are used.

Transition from  the dynamical to the static~\cite{DGLV} approximation in the case of the radiative energy loss is determined through the following two changes and according to the paper~\cite{DynEL1}. The mean free path is altered as:
\begin{eqnarray}
\frac{1}{\lambda_{stat}}= \frac{1}{\lambda_g}{+}\frac{1}{\lambda_q} = 6 \frac{1.202}{\pi^2} \frac{1+n_f/4}{1+n_f/6}\, 3 \alpha_S T{=}c(n_f)\frac{1}{\lambda_{dyn}},\
\label{sd1}
\end{eqnarray}
where $c(n_f = 2.5)\approx 0.84$ is a slowly increasing function of $n_f$ that varies between $c(0)\approx 0.73$ and  $c(\infty)\approx 1.09$ and the effective cross section changes to:
\begin{eqnarray}
v(\mathbf{q})_{stat}=\frac{\mu^2_E}{(\mathbf{q}^2+\mu^2_E)^2} \ . 
\label{sd2}
\end{eqnarray}

\section{Results and Discussion} 
In this section, we concentrate on central 200 GeV Au+Au collisions at RHIC and 2.76 TeV Pb+Pb collisions at the LHC, and investigate how different energy loss ingredients affect the D meson  suppression predictions. Regarding the LHC, for which the high momentum D meson $R_{AA}$ data are available~\cite{D_LHC}, we compare our calculations with experimental data, in order to visually investigate, both qualitative and quantitative, importance of individual effects in  explaining the data.
  
We will start the analysis from the static approximation, which has been historically the first approach to the energy loss calculations. After investigating the adequacy of the static approximation, we will address the importance of including the dynamical scattering centers, the collisional energy loss and the finite size effect. Finally, we will also investigate the importance of finite magnetic mass and the running coupling. 

\begin{figure*}
\epsfig{file=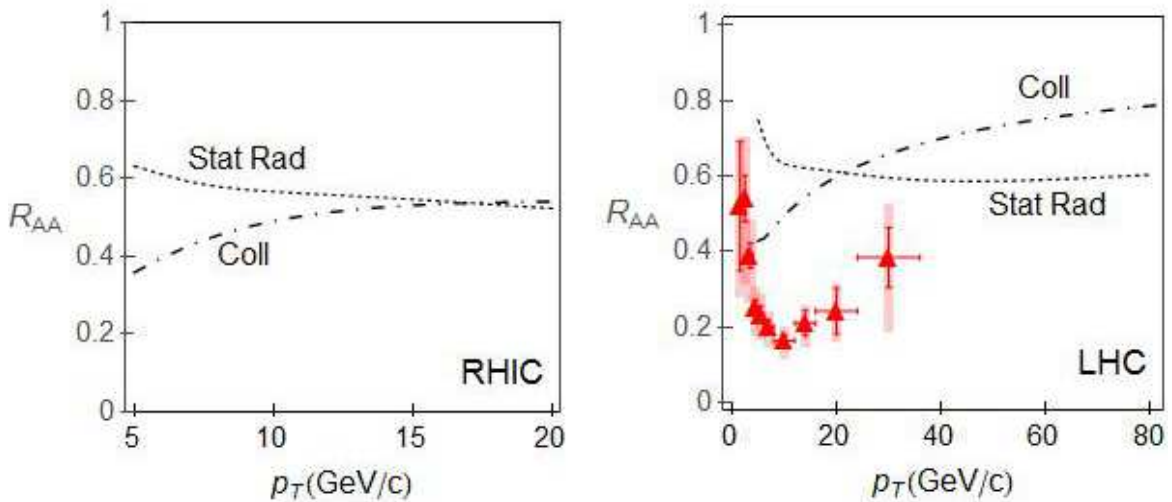,scale=1}
\vspace*{-0.7cm}
\caption{\textbf{Static radiative {\it {\bf vs.}} collisional energy loss suppression.} D meson suppression predictions, as a function of transverse momentum, are shown  for only radiative energy loss in static QCD medium (dotted curve), and for only collisional energy loss in dynamical QCD medium (dot-dashed curve). Left (right) panel corresponds to RHIC (LHC) case. Right panel also shows the D meson $R_{AA}$ data in $0$-$7.5\%$ central 2.76 TeV Pb+Pb collisions at LHC~\cite{D_LHC} (red triangles). Debye mass is $\mu_E=gT$, coupling constant is $\alpha_S=0.3$ ($\alpha_S=0.25$) for RHIC (LHC) and finite magnetic mass effect is not included (i.e. $\mu_M=0$).}
\label{slika1}
\end{figure*}

\begin{figure*}
\epsfig{file=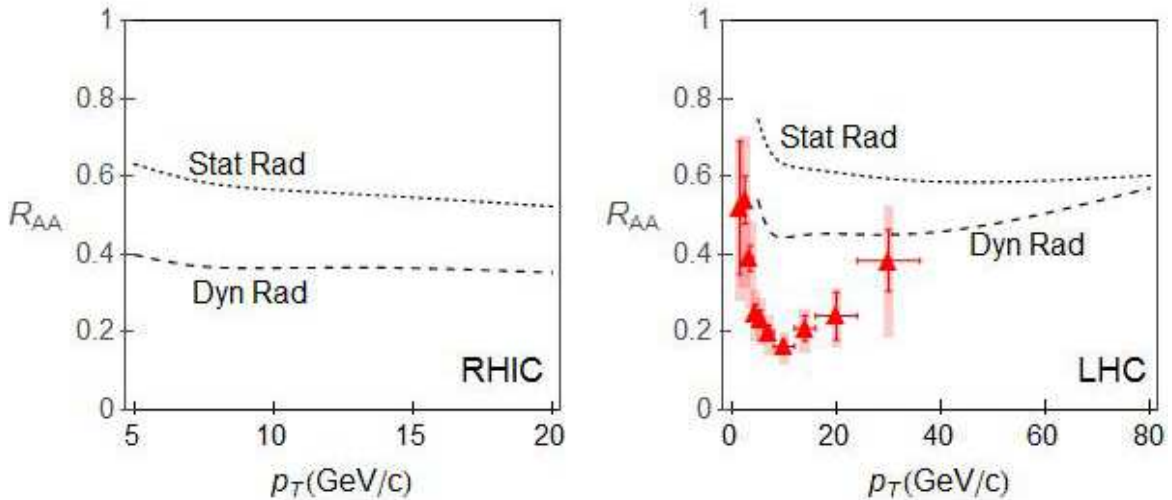}
\vspace*{-0.7cm}
\caption{\textbf{Radiative energy loss suppressions in static {\it {\bf vs.}} dynamical QCD medium.} D meson suppression predictions are shown, as a function of transverse momentum, assuming only radiative energy loss in static (dotted curve) and in dynamical (dashed curve) QCD medium. Left (right) panel corresponds to RHIC (LHC) case. Right panel also shows the D meson $R_{AA}$ data in $0$-$7.5\%$  
central 2.76 TeV Pb+Pb collisions at LHC~\cite{D_LHC} (red triangles). Debye mass is $\mu_E=gT$, coupling constant is $\alpha_S=0.3$ ($\alpha_S=0.25$) for RHIC (LHC) and no finite magnetic mass effect is included (i.e. $\mu_M=0$).}
\label{slika2}
\end{figure*}

\begin{figure*}[h]
\epsfig{file=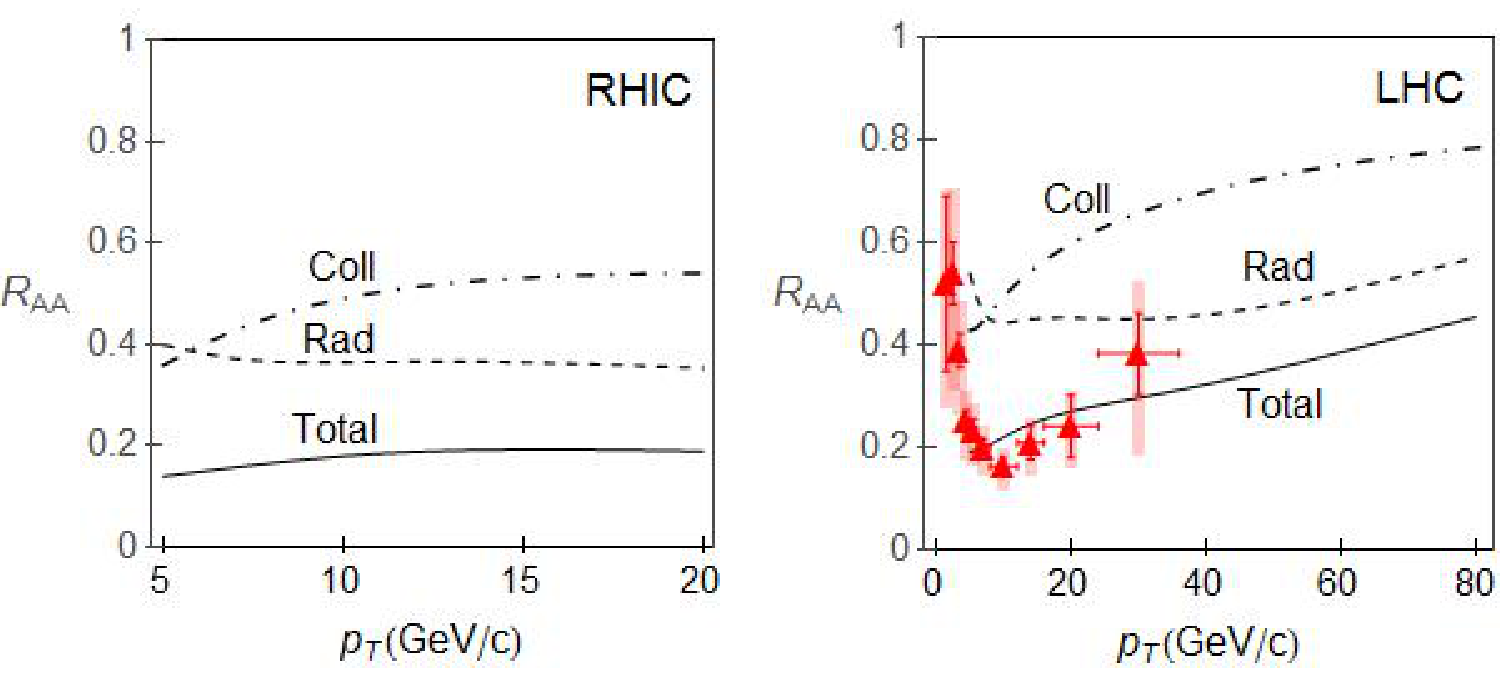}
\vspace*{-0.7cm}
\caption{\textbf{Radiative {\it {\bf vs.}} collisional energy loss suppressions in dynamical QCD medium.} D meson suppression predictions are shown, as a function of transverse momentum, for radiative (dashed curve), collisional (dot-dashed curve) and radiative + collisional (solid curve) energy loss. Left (right) panel corresponds to RHIC (LHC) case. Right panel also shows the D meson $R_{AA}$ data in $0$-$7.5\%$
central 2.76 TeV Pb+Pb collisions at LHC~\cite{D_LHC} (red triangles). Debye mass is $\mu_E=gT$, coupling constant is $\alpha_S=0.3$ ($\alpha_S=0.25$) for RHIC (LHC) and no finite magnetic mass effect is included (i.e. $\mu_M=0$).}
\label{slika3}
\end{figure*}

\begin{figure*}
\epsfig{file=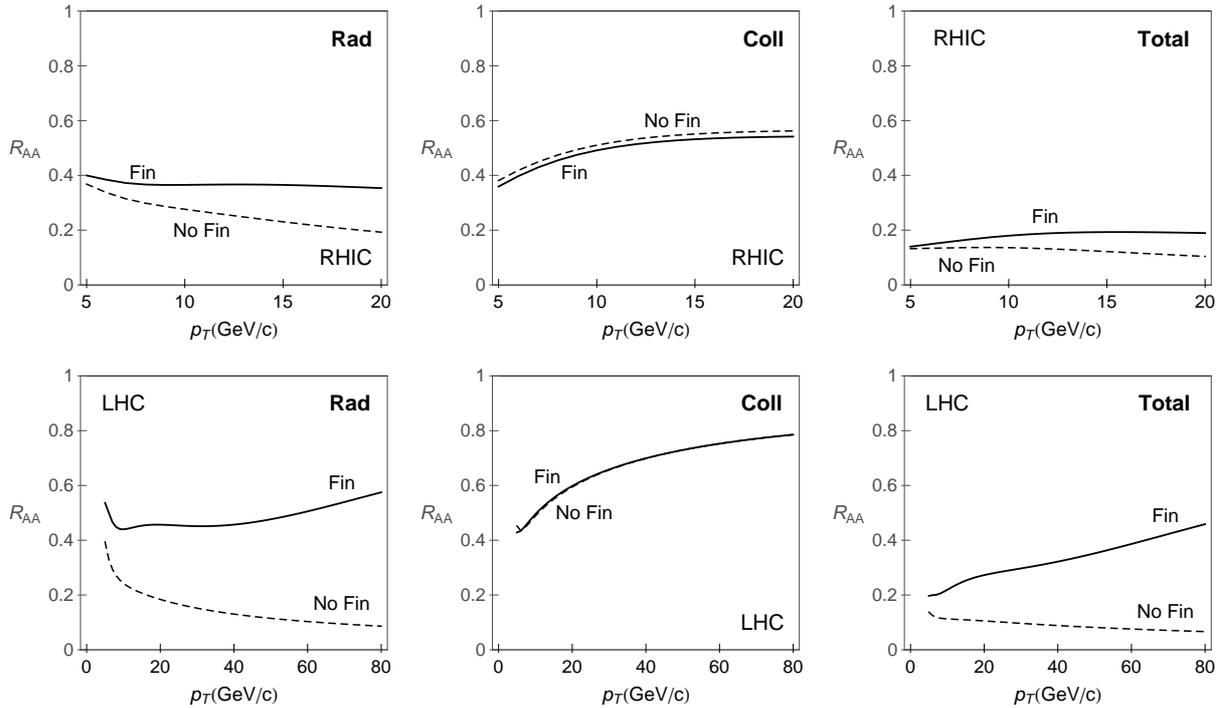,scale=0.45}
\vspace*{-0.5cm}
\caption{\textbf{Finite size effect on $R_{AA}$.} D meson suppression predictions are shown, as a function of transverse momentum, with (solid curve) and without (dashed curve) finite size effect. Upper (lower) panels correspond to RHIC (LHC) case. Left, central and right panel show, respectively, the finite size effect on radiative, collisional and total (radiative + collisional) energy loss in dynamical QCD medium. Debye mass is $\mu_E=gT$, coupling constant is $\alpha_S=0.3$ ($\alpha_S=0.25$) for RHIC (LHC) and no finite magnetic mass effect is included (i.e. $\mu_M=0$).}
\label{slikaLPM}
\end{figure*}

\begin{figure*}
\epsfig{file=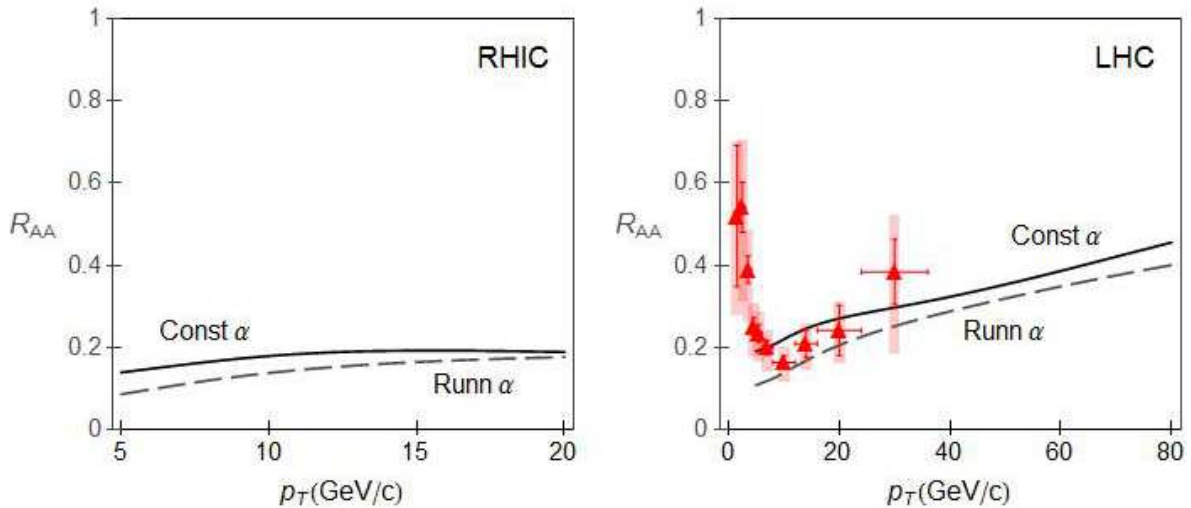}
\vspace*{-0.7cm}
\caption{\textbf{Running coupling effect on $R_{AA}$.} D meson suppression predictions are shown, as a function of transverse momentum, with constant coupling $\alpha_S=0.3$ ($\alpha_S=0.25$) for RHIC (LHC) (solid curve) and with running coupling (dashed curve). No finite magnetic mass effect is included (i.e. $\mu_M=0$). In both cases radiative + collisional contributions in dynamical QCD medium are included. Left (right) panel corresponds to RHIC (LHC) case. Right panel also shows the D meson $R_{AA}$ data in $0$-$7.5\%$
central 2.76 TeV Pb+Pb collisions at LHC~\cite{D_LHC} (red triangles). }
\label{slika4}
\end{figure*}

\begin{figure*}
\epsfig{file=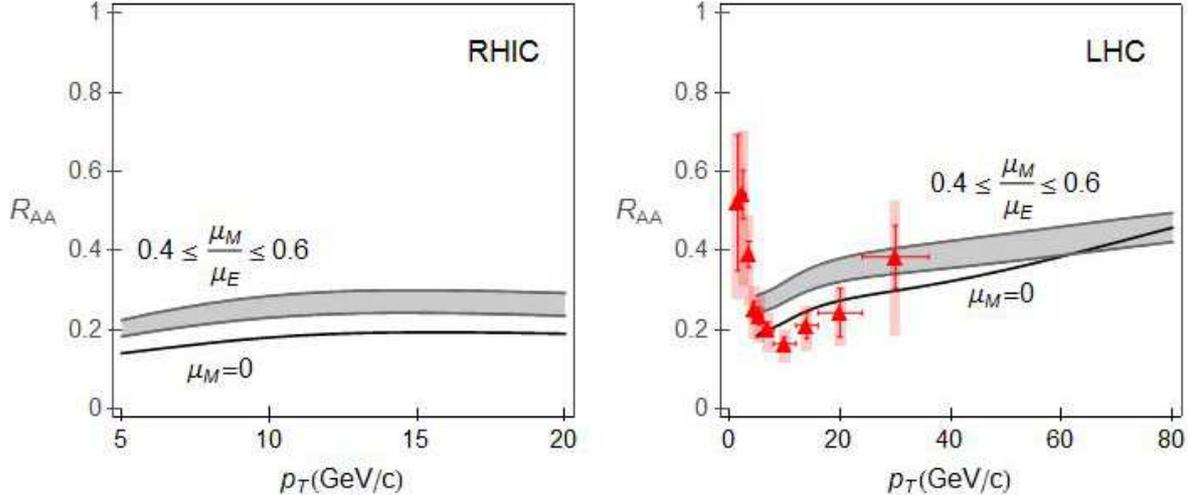}
\vspace*{-0.7cm}
\caption{\textbf{Magnetic mass effect on $R_{AA}$.} D meson suppression predictions are shown, as a function of transverse momentum, for radiative + collisional energy loss in dynamical QCD medium, with (gray band) and without (solid curve) magnetic mass. Left (right) panel corresponds to RHIC (LHC) case. Right panel also shows the D meson $R_{AA}$ data in $0$-$7.5\%$
central 2.76 TeV Pb+Pb collisions at LHC~\cite{D_LHC} (red triangles). Debye mass is $\mu_E=gT$ and coupling constant is $\alpha_S=0.3$ ($\alpha_S=0.25$) for RHIC (LHC). Upper (lower) boundary of each band corresponds to $\mu_M/\mu_E=0.6$ ($\mu_M/\mu_E=0.4$).}
\label{slika5}
\end{figure*}

\begin{figure*}
\epsfig{file=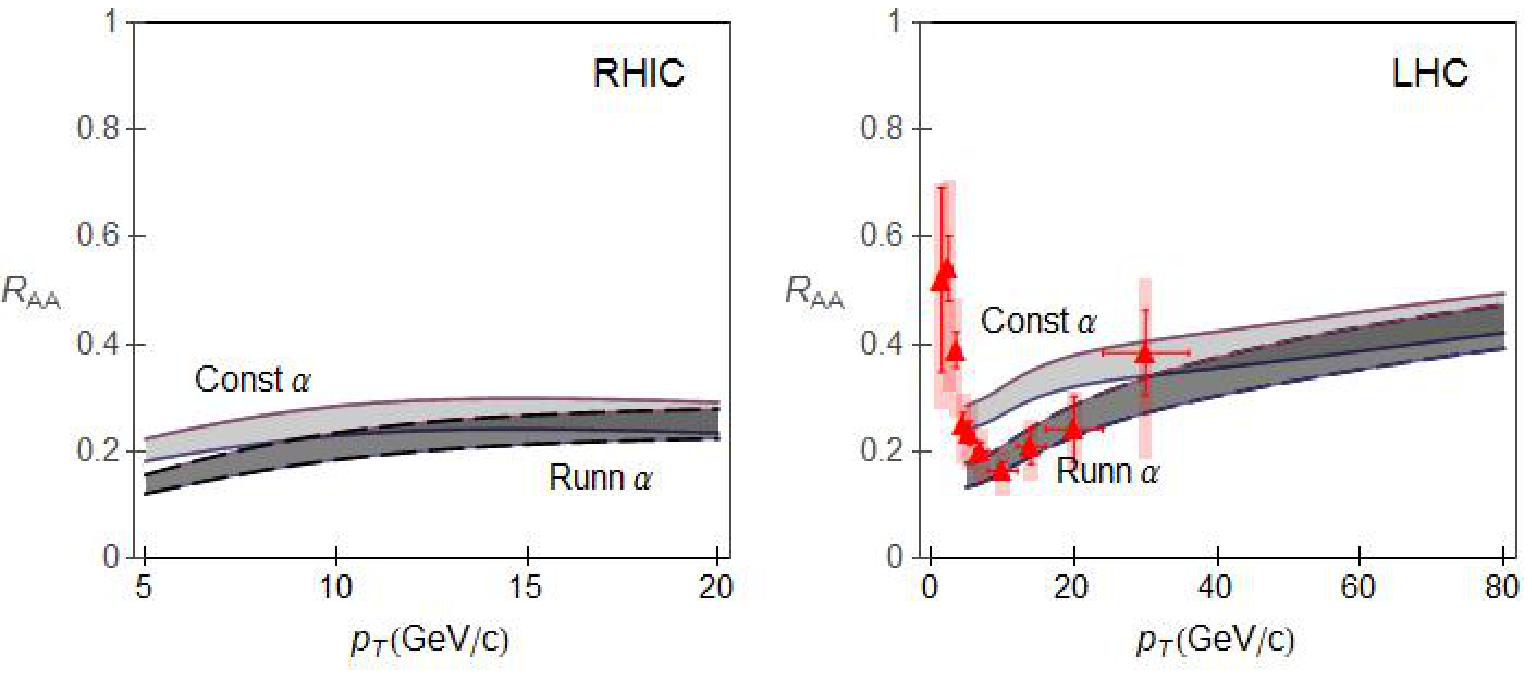}
\vspace*{-0.7cm}
\caption{\textbf{Running coupling and magnetic mass effect on $R_{AA}$.} D meson suppression predictions are shown, as a function of transverse momentum, with the constant coupling $\alpha_S=0.3$ ($\alpha_S=0.25$) for RHIC (LHC) (light gray band) and with the running coupling (dark gray band). In both cases radiative + collisional contributions in dynamical QCD medium are included. Upper (lower) boundary of each band corresponds to $\mu_M/\mu_E=0.6$ ($\mu_M/\mu_E=0.4$). Left (right) panel corresponds to RHIC (LHC) case. Right panel also shows the D meson $R_{AA}$ data in $0$-$7.5\%$
central 2.76 TeV Pb+Pb collisions at LHC~\cite{D_LHC} (red triangles). }
\label{slika6}
\end{figure*}

We therefore start from the static approximation, where we use a fixed value of the strong coupling constant $\alpha_S=\frac{g^2}{4 \pi}=0.3$ at RHIC ($\alpha_S=0.25$ at LHC) and Debye screening mass $\mu_E \approx g T$. Note that these values are used in Figs.~\ref{slika1}-\ref{slikaLPM} and Fig.~\ref{slika5}. Also, note that magnetic mass effect is not included ($\mu_M=0$) in Figs.~\ref{slika1}-\ref{slika4}, while the finite magnetic mass is considered in Figs.~\ref{slika5} and~\ref{slika6}. The running coupling is considered in Figs.~\ref{slika4} and~\ref{slika6}. The finite size QCD medium is considered in each figure, whereas Fig.~\ref{slikaLPM} investigates the significance of the finite size effect.

To test the adequacy of the widely used static approximation (modeled by Yukawa potential)~\cite{GW}, we compare relative importance of radiative and collisional energy loss contributions to the suppression predictions. Namely, in the static approximation, collisional energy loss has to be equal to zero, i.e. the static approximation implies that collisional energy loss can be neglected compared to radiative energy loss. However, in Fig.~\ref{slika1}, we see that the suppression due to collisional energy loss is comparable - or even larger - compared to the radiative energy loss suppression.  

This then clearly shows that the static approximation is not an adequate one for the D meson suppression calculations, and that the collisional energy loss has to be taken into account in the suppression predictions. Therefore, a number of the approaches which take only radiative energy loss (for an overview see~\cite{RELO}) - and some that take only collisional energy loss (e.g.~\cite{CEL, CEL1}) - are clearly not adequate. This can also be directly observed in the right panel of Fig.~\ref{slika1}, where we see that the static approximation leads to a strong disagreement with the data, i.e. to 2-3 times smaller suppression than the one experimentally observed. Consequently, we will below first test the importance of including the dynamical effects in radiative energy loss (Fig.~\ref{slika2}) and then also test the importance of collisional energy loss within such dynamical medium (Fig.~\ref{slika3}). 

Therefore, in Fig.~\ref{slika2}, we compare the D meson suppression obtained only from radiative energy loss in the static framework, with the one in the dynamical framework. We observe a large difference in the two suppressions, with a significant suppression increase in the dynamical case. Consequently, the dynamical energy loss effect has to be taken into account at RHIC, as there are no momenta within RHIC jet momentum range where static approximation becomes adequate.  At the LHC, the results indicate that, for jet momentum ranges larger than 100~GeV/c, the static approximation to {\it radiative} energy loss may become valid, in general agreement with~\cite{BDMPS,BDMS,DynEL,DynEL1}; note however that the dynamical effect has to be taken into account even for these momenta, as the collisional energy loss - which is zero in the static approximation - gives a significant contribution to the jet suppression (see the right panel in Fig.~\ref{slika1}).   However, despite the fact that inclusion of dynamical effect significantly increases the suppression compared to the static approximation, from the right panel in Fig.~\ref{slika2} we see that, at least below 50 GeV/c, radiative energy loss alone is not able to neither quantitatively nor qualitatively (see the shape of the curve) explain the experimental data, which leads to the conclusion that including only radiative energy loss to model the jet-medium interaction is clearly not adequate.

Furthermore, the results shown in Fig.~\ref{slika2} imply a question whether a collisional energy loss is still relevant in the dynamical QCD medium, as suppression due to radiative energy loss significantly increases in the dynamical QCD medium. To address this question, in Fig.~\ref{slika3} we compare the D meson suppressions resulting from collisional and  radiative energy loss, both calculated in the dynamical QCD medium. We observe that, even when the dynamical effect is accounted, suppressions from both radiative and collisional contributions are important (consistently with claims in Refs.~\cite{CR,CR1,MD_Coll}). This further underscores that collisional energy loss has to be included in the D meson suppression predictions at both RHIC and the LHC. Moreover, we see that including the collisional contribution increases D meson suppression by up to factor of two comparing to the suppression resulting only from dynamical radiative energy loss. Consistently with this observation, we see  that the total suppression is significantly larger than either of the two contributions - radiative alone or collisional alone - so that they jointly have to be taken into account for the accurate predictions. Furthermore, our main observation from Fig.~\ref{slika3} is that inclusion of the dynamical effect results in a (rough) agreement with the experimental data, which leads to the conclusion that the dynamical effect is the main/necessary ingredient for accurate description of the jet-medium interactions.
 
Since we showed that collisional and radiative energy losses are important, we will further investigate how they are affected by the finite size (LPM) effect, as it is commonly considered that this effect is not important for heavy flavor at RHIC. In Fig.~\ref{slikaLPM}, we separately investigate the finite size effect for radiative (the left panels), collisional (the central panels) and radiative plus collisional (the right panels) energy loss; the top and the bottom panels correspond to the RHIC and the LHC cases, respectively.

We see that for D mesons at both RHIC and the LHC, the finite size effect is indeed negligible for collisional energy loss, but that they are significant for both radiative and total energy loss suppressions. That is, we see that neglecting LPM effect can lead to as much as two times larger suppression at RHIC and several times larger suppression at the LHC. Also, LPM effect leads to qualitatively different suppression dependence on momenta, as this effect can lead to a decrease - rather than increase - of suppression with jet momentum.  Consequently, LPM effect has to be taken into account in heavy flavor suppression predictions at both RHIC and the LHC.  
 
\setlength\belowcaptionskip{-3ex}
We next consider how the running coupling~\cite{RunnC} affects the $R_{AA}$. Therefore, in Fig.~\ref{slika4} we compare the D meson suppression predictions obtained by using the fixed value of strong coupling constant, with the predictions when the running coupling is accounted, as a function of the transverse momentum. From Fig.~\ref{slika4} we observe that the running coupling leads to an increase in the suppression by almost a factor of two at lower jet momenta, while it makes no significant difference at higher jet momenta. Note that such an unequal contribution notably changes the shape of the suppression pattern, so that accounting for the running coupling for D mesons at both RHIC and the LHC, is also important. Furthermore, when comparing the predictions with available (LHC) experimental data (see the right panel of Fig.~\ref{slika4}), we see that inclusion of running coupling leads to a somewhat worse agreement with experimental data, compared to the predictions with constant coupling; we will however see below that inclusion of {\it both} the running coupling and the finite magnetic mass improves the predictions.

We next investigate the significance of taking into account the finite magnetic mass in the suppression calculations. 
Namely, all previous energy loss calculations assumed zero magnetic mass, in accordance with the perturbative QCD. However, different non-perturbative approaches~\cite{xb, xb4, xb1, xb2, xb3} reported a non-zero magnetic mass at RHIC and the LHC, which indicates that the finite magnetic mass has to be included in the radiative energy loss calculations~\cite{MagM}.
 
\setlength\belowcaptionskip{-3ex}

Hence in Fig.~\ref{slika5} we compare the D meson suppression predictions with and without the finite magnetic mass included in the suppression calculations. To investigate the importance of magnetic mass only, we do not include running coupling in this figure, i.e. we assume the constant coupling. Figure~\ref{slika5} shows that the inclusion of the finite magnetic mass effect leads to a notable $\sim$ 30\% decrease in the suppression. Consequently, the finite magnetic mass effect is also important. Furthermore, when comparing the predictions with available (LHC) experimental data (see the right panel of Fig.~\ref{slika5}), we see that the effect of the inclusion of magnetic mass runs in the opposite direction from the inclusion of running coupling, and also in itself leads to a worse agreement with experimental data (compared to predictions with zero magnetic mass). From this and the previous figure (i.e. Figs.~\ref{slika4} and~\ref{slika5}), one can conclude that inclusion of the individual improvements in the energy loss calculations - in particular the running coupling alone, or the magnetic mass alone - does not necessarily lead to the improvement in the agreement between the predictions and the data. 

Consequently, we finally consider how the inclusion of both the running coupling~\cite{RunnC} and the magnetic mass affects $R_{AA}$. Therefore, in Fig.~\ref{slika6} we use the finite value of magnetic mass, and compare the D meson suppression predictions with fixed value of strong coupling constant, with those when the running coupling is used, as a function of transverse momentum.  We see that these two effects, taken together, lead to a very good agreement with the experimental data, i.e. to both quantitative and qualitative improvement compared to the case in Fig.~\ref{slika3}. This illustrates possible synergy in including different energy loss effects: taken individually the running coupling and the finite magnetic mass lead to worse agreement with the experimental data, but taken together they notably improve the agreement. Therefore, detailed study of parton energy loss, as well as inclusion of all important medium effects may be necessary to correctly model the interactions of high-momentum particles with the QCD medium.

\section{Conclusions} 

Since our dynamical energy loss formalism led to a robust  agreement with the experimentally measured nuclear modification factor for different experiments, probes and experimental conditions (i.e. centrality ranges)~\cite{RunnC,NCLHC, CRHIC}, we investigated how different energy loss effects contribute to such a good agreement. In particular, we aimed determining whether such a good agreement is a consequence of a single dominant effect, or  of several smaller improvements. We here investigated this issue for the case of D mesons, whose suppression patterns are not modified by the fragmentation functions, so that they present a clear energy loss probe. We used an approach where we started from the simplest reasonable (and historically justified) energy loss model - which includes only radiative energy loss - and then sequentially added different model improvements. This approach both allows investigating the importance of different energy loss ingredients, and obtaining the historical perspective on how the energy loss model has been improved. In particular, we studied the importance of the transition from the static to the dynamical framework and of including collisional energy loss, the finite size effect, the finite magnetic mass and the running coupling. As an overall conclusion, we found that the most important effect in modeling the jet-medium interactions is inclusion of the dynamical effect, i.e. modeling the medium constituents as dynamical (moving) particles, instead of the commonly used static scattering centers. However, for a fine agreement with the data, we find that each energy loss effect is important, and that the robust agreement between the theoretical predictions and the experimental data is a cumulative effect of all these improvements.  As an outlook, the presented results suggest that further improvements in the energy loss model may be significant for accurately explaining the data even outside of the energy ranges and observables that we tested so-far. Therefore we expect that data from the upcoming RHIC and LHC runs will help testing - or even further constraining - model calculation at higher transverse momentum.

{\em Acknowledgments:} 
This work is supported by Marie Curie International Reintegration Grant 
within the $7^{th}$ European Community Framework Programme 
PIRG08-GA-2010-276913 and by the Ministry of Science and Technological 
Development of the Republic of Serbia, under project No. ON171004.

\end{document}